\documentclass[10pt]{amsart}
\usepackage{amsmath,amssymb,graphicx}

\begin{document}

\newtheorem{thm}{Theorem}[section]
\newtheorem{lem}[thm]{Lemma}
\newtheorem{prop}[thm]{Proposition}
\newtheorem{cor}[thm]{Corollary}
\newtheorem{conj}[thm]{Conjecture}
\newtheorem{defn}[thm]{Definition}
\newtheorem*{remark}{Remark}

\numberwithin{equation}{section}

\newcommand{\Z}{{\mathbb Z}} 
\newcommand{\Q}{{\mathbb Q}}
\newcommand{\R}{{\mathbb R}}
\newcommand{\C}{{\mathbb C}}
\newcommand{\N}{{\mathbb N}}
\newcommand{\FF}{{\mathbb F}}
\newcommand{\T}{{\mathbb T}}
\newcommand{\fq}{\mathbb{F}_q}
\newcommand{\PP}{{\mathbb P}}
\newcommand{\E}{{\mathbb E}}

\def\scrA{{\mathcal A}}
\def\scrB{{\mathcal B}}
\def\scrD{{\mathcal D}}
\def\scrE{{\mathcal E}}
\def\scrH{{\mathcal H}}
\def\scrI{{\mathcal I}}
\def\scrJ{{\mathcal J}}
\def\scrK{{\mathcal K}}
\def\scrL{{\mathcal L}}
\def\scrM{{\mathcal M}}
\def\scrN{{\mathcal N}}
\def\scrQ{{\mathcal Q}}
\def\scrS{{\mathcal S}}

\newcommand{\rmk}[1]{\footnote{{\bf Comment:} #1}}

\renewcommand{\mod}{\;\operatorname{mod}}
\newcommand{\ord}{\operatorname{ord}}
\newcommand{\TT}{\mathbb{T}}
\renewcommand{\i}{{\mathrm{i}}}
\renewcommand{\d}{{\mathrm{d}}}
\renewcommand{\^}{\widehat}
\newcommand{\HH}{\mathbb H}
\newcommand{\G}{\mathbb G}
\newcommand{\Vol}{\operatorname{vol}}
\newcommand{\area}{\operatorname{area}}
\newcommand{\tr}{\operatorname{tr}}
\newcommand{\norm}{\mathcal N} 
\newcommand{\intinf}{\int_{-\infty}^\infty}
\newcommand{\ave}[1]{\left\langle#1\right\rangle} 
\newcommand{\Var}{\operatorname{Var}}
\newcommand{\Cov}{\operatorname{Cov}}
\newcommand{\Prob}{\operatorname{Prob}}
\newcommand{\sym}{\operatorname{Sym}}
\newcommand{\disc}{\operatorname{disc}}
\newcommand{\CA}{{\mathcal C}_A}
\newcommand{\cond}{\operatorname{cond}} 
\newcommand{\lcm}{\operatorname{lcm}}
\newcommand{\Kl}{\operatorname{Kl}} 
\newcommand{\leg}[2]{\left( \frac{#1}{#2} \right)}  
\newcommand{\SL}{\operatorname{SL}}
\newcommand{\interior}{\operatorname{int}}
\newcommand{\sgn}{\operatorname{sgn}}
\newcommand{\Id}{\operatorname{Id}}

\newcommand{\be}{\begin{equation}}
\newcommand{\ee}{\end{equation}}
\newcommand{\bs}{\begin{split}}
\newcommand{\es}{\end{split}}
\newcommand{\bra}{\left\langle}
\newcommand{\ket}{\right\rangle}

\newcommand{\sumstar}{\sideset \and^{*} \to \sum}

\newcommand{\LL}{\mathcal L} 
\newcommand{\sumf}{\sum^\flat}
\newcommand{\Hgev}{\mathcal H_{2g+2,q}}
\newcommand{\USp}{\operatorname{USp}}
\newcommand{\conv}{*}
\newcommand{\dist} {\operatorname{dist}}
\newcommand{\CF}{c_0} 
\newcommand{\kerp}{\mathcal K}

\newcommand{\gp}{\operatorname{gp}}
\newcommand{\Area}{\operatorname{Area}}
\newcommand{\diam}{\operatorname{diam}}
\newcommand{\supp}{\operatorname{supp}}

\title[Flat tori with random impurities]{Spectral geometry of flat tori \\ with random impurities}

\author{Henrik Uebersch\"ar}
\address{Max Planck Institute of Mathematics, Vivatsgasse 7, 53111 Bonn, Germany 
\and 
Laboratoire Paul Painlev\'e, Universit\'e Lille 1, 59655 Villeneuve d'Ascq, France.}
\email{uebersch@mpim-bonn.mpg.de}
\date{\today}

\maketitle

\begin{abstract}
We discuss new results on the geometry of eigenfunctions in disordered systems. More precisely, we study tori $\R^d/L\Z^d$, $d=2,3$, with uniformly distributed Dirac masses. Whereas at the bottom of the spectrum eigenfunctions are known to be localized, we show that for sufficiently large eigenvalue there exist uniformly distributed eigenfunctions with positive probability. We also study the limit $L\to\infty$ with a positive density of random Dirac masses, and deduce a lower polynomial bound for the localization length in terms of the eigenvalue for Poisson distributed Dirac masses on $\R^d$. Finally, we discuss some results on the breakdown of localization in random displacement models above a certain energy threshold.
\end{abstract}

\section{Introduction}

In 1900 Paul Drude introduced a classical model \cite{Drude,Drude2} for the motion of electrons in a material with the aim of studying transport properties such as conductivity of metals. In particular Ohm's law could be derived from the model. In 1933 the model was supplemented with results from quantum theory by Hans Bethe and Arnold Sommerfeld as what is now known as the Drude-Sommerfeld or free electron model.

Anderson discovered in 1958 that at low energy electronic transport could break down in disordered media, provided the disorder is sufficiently strong \cite{Anderson}. In particular the eigenstates in this regime are exponentially localized. This phenomenon is today known as Anderson localization. A key question concerns the transition from a localized to a delocalized regime when the disorder becomes small compared with the energy. In 1979, Abrahams, Anderson, Licciardello and Ramakrishnan proposed their scaling theory which suggests that such a transition should exist in dimension $d\geq 3$, the case $d=2$ being critical, whereas for $d=1$ there is only a localized regime. Regarding $d=2$ the widely held belief is that no transition exists, however the localization length can be very large compared with the size of the system.

The type of disordered system considered in this paper is a box $\Lambda=[L/2,L/2]^d$, $L>0$, which contains $N$ independently uniformly distributed impurities, which are modeled by Dirac masses (also known as Fermi's pseudo-potential, delta potentials or point scatterers). In this paper we will consider periodic boundary conditions, i.e. we will study flat tori $\T^d_L=\R^d/L\Z^d$. However, our results can easily be generalized for Dirichlet or Neumann boundary conditions.

\section{Tori with random impurities}

By a random Schr\"odinger operator on $\T^d_L$ we mean the following type of stochastic differential operator
\begin{equation}
H_\Omega=-\Delta+\sum_{\omega\in\Omega}V(x-\omega), \quad V\in C^0(\T^d_L),
\end{equation}
where $\Omega$, $|\Omega|=N$, is a stochastic process on $\T^d_L$.

A simplified model $H$ replaces $V=\delta$, where $\delta$ denotes a Dirac mass centered at the origin. Such a ``potential'' is known as Fermi's pseudo-potential, a delta potential or a point scatterer. The model approximates the generic random Schr\"odinger operator $H_\Omega$ well in the regime, where $\diam\supp V\ll w(E)$, and where $w(E)\asymp 1/\sqrt{E}$ denotes the wavelength, and $E$ the energy.

The advantage of considering the formal operator $H$ is that it can be realized as a self-adjoint extension of the restricted Laplacian $H_0=-\Delta|_{C^\infty_c(\T^d_L-\Omega)}$, which is a positive symmetric operator with deficiency indices $(N,N)$. The extensions are parametrized by a unitary matrix $U\in U(N)$, and the corresponding self-adjoint extension is given by the restriction of the adjoint $H_0^*$ to the domain of functions $f\in H^2(\T^d_L-\Omega)$ which are of the form 
$$f(x)=g(x)+\left\langle v,\G_{\i}(x)\right\rangle+\left\langle Uv,\G_{-\i}(x)\right\rangle$$
and $\G_{\pm\i}(x)=(G_{\pm\i}(x,\omega_1),\cdots,G_{\pm\i}(x,\omega_N))$, where $G_\i(x,\omega)$ and $G_{-\i}(x,\omega)$ are incoming and outgoing circular waves at the point $\omega\in\T^d_L$.

We note that the parameter space of self-adjoint extensions is much larger than the physical parameter space which is of dimension $N$ (there are $N$ real coupling constants). In fact a non-diagonal matrix $U$ corresponds to an extension which violates local conservation of mass in the scattering process (i. e. it is possible for part of the wave to enter into one impurity and emerge from another, as if a wire were attached between the two points). Assuming local conservation of mass in all scattering processes we reduce the parameter space to the subgroup of diagonal unitary matrices $D(N)\subset U(N)$. Note that the operator $H$ corresponds to a matrix $D=e^{\i\varphi}\Id$, for some $\varphi\in(-\pi,\pi)$, since all coupling constants are equal to $1$.

\section{Anderson localization}

\begin{figure}
\includegraphics[scale=.8]{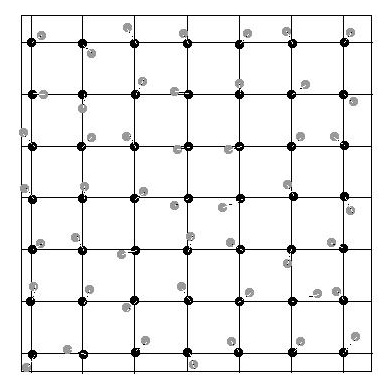}
\caption{The picture above shows a random displacement model in two dimensions. In black the lattice $\Z^2$, in grey the randomly perturbed lattice $\Z^2+\omega$.}
\end{figure}

Let us consider a particular type of random Schr\"odinger operator on $\R^d$, a random displacement model (cf. Figure 1)
\begin{equation}
-\Delta+\sum_{\xi\in\Z^d}V(x-\xi-\omega_\xi), \quad V\in C_c^\infty(\R^d)
\end{equation}
where the displacements $\omega_\xi$ are i.i.d. random variables with radially symmetric probability density $\PP(x)=P(|x|)$, $P(0)>0$ and $P\in C^\infty_c(\R_+)$ strictly decreasing. The disorder in the present system can for instance be measured by the average displacement, $\delta_0=\E_\PP(X)$, where $X$ is a random variable with probability density $\PP$.

If the probability distribution $\PP$ were given by a Dirac distribution centered at the origin, i.e. if $\Z^d+\omega=\Z^d$ and $\delta_0=0$, then we know from Bloch's theorem that the spectrum has a continuous band structure with corresponding extended eigenstates. However, if there is sufficiently strong disorder, i.e. $\delta_0$ is large enough, then we have, almost surely, pure point spectrum and the corresponding eigenstates are exponentially localized at the bottom of the spectrum.

\begin{defn}\label{loc}
Let $E_2>E_1 >0$. We say that $H_\Omega$ is exponentially localized on $[E_1,E_2]$, if $H_\Omega$ has a.s. pure point spectrum on $[E_1,E_2]$ and the eigenfunctions $\psi_\lambda$ satisfy the
bound :
$$\exists x_\lambda\in\R^d: \forall x\in\R^d: \quad |\psi_\lambda(x)|\leq Ce^{|x-x_\lambda|/L_{loc}}$$
where the localization length $L_{loc}$ depends on $E_1,E_2$ and the localization
centre $x_\lambda$ as well as the constant $C$ depend on the sample $\omega\in\Omega$.
\end{defn}

Mathematical proofs of Anderson localization were for instance obtained by Goldsheid, Molchanov and Pastur \cite{Pastur}, Fr\"ohlich and Spencer \cite{FroehlichSpencer}, Simon \cite{Simon}, Bourgain and Kenig \cite{BourgainKenig}, Klopp \cite{Klopp} and Germinet \cite{Germinet}. However, a crucial question concerns the existence of a delocalization transition in the regime where the disorder is small compared with the energy, in the case of a random displacement model: $\delta_0\ll E$. 

In 1979, the ``Gang of Four'', Abrahams, Anderson, Licciardello and Ramakrishnan \cite{AALR} proposed what is now known as the ``scaling theory of localization'' which predicted that the answer should depend on the dimension $d$. For $d=1$ the theory predicts that there is only a localized regime, i.e. any amount of disorder is sufficient to localize all eigenstates. For $d\geq3$ it predicts the existence of a delocalization transition, i.e. when the disorder is sufficiently small compared with the energy, the eigenstates should become delocalized. The case $d=2$ was identified as critical, and it was famously conjectured that there should only be a localized regime, however, the localization length could be very large.

The mathematical characteristics of the delocalized regime are continuous band structure above a certain energy threshold and associated extended eigenstates. A very interesting phenomenon is observed near the transition point between the two regimes, where physicists have observed a multifractal structure of the eigenstates \cite{Huckestein}. 

Very few rigorous mathematical results exist regarding the multifractal or delocalized regime. We will continue to describe some rigorous results about the geometry of the eigenstates of Poisson random Schr\"odinger operators with Dirac masses on flat tori. In particular these compact models allow us to give polynomial lower bounds on the localization length on $\R^d$. In fact for certain stochastic processes, such as random displacement models, one can go much further and prove the breakdown of exponential localization for sufficiently large eigenvalues \cite{U2}. We will sketch these results at the end of this paper.

\section{Spectrum and eigenfunctions}

We consider a flat torus $\T^d_L$, $d=2,3$ with $N$ independent uniformly distributed delta potentials located at the points $\Omega=\{x_1,\cdots,x_N\}$and we let $U=e^{\i\varphi}\Id_N$. We denote the associated self-adjoint extension of $-\Delta|_{C^\infty_c(\T^d_L-\Omega)}$ by $-\Delta_\varphi$. We denote the Green's function, the resolvent kernel of the Laplacian on $\T^d_L$ by $$G_\lambda(x,y)=\frac{1}{-\Delta-\lambda}\delta(x-y), \quad \lambda\notin\sigma(-\Delta).$$

The spectrum of $-\Delta_\varphi$ consists of two parts, old eigenfunctions of the Laplacian which vanish on $\Omega$ and new eigenfunctions which, almost surely, diverge at each point in $\Omega$. The divergence is of order $\log|x-x_j|$ in $d=2$ and of order $1/|x-x_j|$ in $d=3$.

The new eigenvalues are solutions of the equation 
\begin{equation}
\det A_\lambda^\varphi=0
\end{equation}
where the matrix entries are given by 
$$(A^\varphi_\lambda)_{kl}=G_\lambda(x_k,x_l)-\Re G_\i(x_k,x_l)-\tan(\frac{\varphi}{2})\Im G_\i(x_k,x_l).$$
In particular one can show that per old Laplacian eigenspace the self-adjoint extension $-\Delta_\varphi$ produces at most $N$ new eigenvalues, which almost surely lie in between two neighbouring Laplacian eigenvalues.

The associated new eigenfunctions are superpositions of Green's functions
\begin{equation}
\psi_\lambda(x)=\left\langle v,\G_\lambda(x)\right\rangle, \quad v\in\ker A_\lambda^\varphi
\end{equation}
and we recall $\G_\lambda(x)=(G_\lambda(x,x_1),\cdots,G_\lambda(x,x_N))$. Also note that $v$ is a function of the random variables $x_1,\cdots,x_N$. Given $\det A_\lambda^\varphi=0$, the matrix $A_\lambda^\varphi$ has almost surely rank $N-1$ which implies $\dim\ker A_\lambda^\varphi=1$, so the choice of $v$ is unique up to normalization.

\section{Uniformly distributed eigenfunctions}

The aim of this section is to given information on the geometry of the new eigenfunctions of the random operator $-\Delta_\varphi$. For a large number $N$ of impurities we expect the eigenfunctions to be exponentially localized at the bottom of the spectrum. However, we will show that for sufficiently large eigenvalues there exist, with a certain positive probability (which tends to zero as $N\to\infty$), eigenfunctions which are uniformly distributed on the torus $\T^d_L$. In particular this fact gives information on the size of the localization length, yielding a certain polynomial lower bound.

Before we state our main result we recall that each new eigenvalue $\lambda$ satisfies almost surely $\lambda\in(\lambda_j,\lambda_{j+1})$ for some $j$.
We have the following result.
\begin{thm}
Denote the Laplacian eigenvalues on $\T^d_L$ by $\{\lambda_j\}_{j\in\N}$. Fix any $a_0\in C^\infty(\T^d_L)$. There exists $\scrJ_d=\{j_k\}_k\subset\N$, an index subset of full density such that for any $j\in\scrJ_d$ and any new eigenvalue of $-\Delta_\varphi$, $\lambda\in(\lambda_j,\lambda_{j+1})$, we have with probability $\gtrsim\frac{1}{N}$ for all $a\in C^\infty(\T^d_L)$, s.t. $|\hat{a}|\leq|\widehat{a_0}|$\footnote{By which we mean $\forall\zeta\in\Z^d: |\hat{a}(\zeta)|\leq|\widehat{a_0}(\zeta)|$.},
\begin{equation}\label{equidist}
\int_{\T^d_L}a(x)|\psi_\lambda(x)|^2dx=\frac{1}{L^d}\int_{\T^d_L}a(x)dx+O_\epsilon(N^{1/2}\|\widehat{a_0}\|_{l^1}\lambda^{-\delta_d+\epsilon}L^{-2\delta_d+\epsilon})
\end{equation}
and $\delta_2=\tfrac{17}{416}$, $\delta_3=\frac{1}{12}$.
\end{thm}

We may now apply this result to give a lower bound on the localization length. The equidistribution theorem above implies that if we are still in the localized regime, then the localization length must exceed the size of the torus: $$L_{loc}\gg L.$$

Let us now fix a positive density of impurities $$N=\rho L^d$$ and observe that our stochastic process on $\T^d_L$ converges to a Poisson process of density $\rho$ on $\R^d$ in the limit $L\to\infty$. Our equidistribution result on the torus of $\T^d_L$ therefore gives information on the localization length for a Schr\"odinger operator with Poisson delta potentials on $\R^d$. 

In particular, equidistribution occurs when
$$N^{1/2}\lambda^{-\delta_d}L^{-2\delta_d}\ll L^{-d} \Leftrightarrow L\ll\lambda^{\alpha_d}$$
where
$$\alpha_d=\frac{\delta_d}{\tfrac{3d}{2}+2\delta_d}.$$

And this implies that the localization length must satisfy the lower bound $$L_{loc}\gtrsim \lambda^{\alpha_d}.$$

\section{Elements of the proof}

Consider the square torus $\T^2=\R^2/2\pi\Z^2$.

Let $$\Psi(x)=\sum_{j=1}^N v_j G_\lambda(x,x_j).$$
We introduce a test function $a\in C^\infty(\T^2)$ and the $L^2$-normalized eienfunction is denoted by $\psi_\lambda=\Psi_\lambda/\|\Psi_\lambda\|_2$.

We have $$\left\langle a \psi_\lambda,\psi_\lambda\right\rangle=\frac{1}{4\pi^2}\int_{\T^2} a d\mu +\sum_{\zeta\in\Z^2} \hat{a}(\zeta)\left\langle e_\zeta \psi_\lambda,\psi_\lambda\right\rangle$$
were $$e_\zeta(x)=\frac{1}{2\pi}e^{\i \left\langle\zeta,x\right\rangle}.$$

Our goal is to show that a matrix element correponding to a nonzero mode $\left\langle e_\zeta \psi_\lambda,\psi_\lambda\right\rangle$, $\zeta\neq0$, is small if $\lambda\gg 1$.

Let us define the annulus $$A_\zeta(\lambda,L)=\{\xi\in\Z^2 \mid ||\xi-\zeta|^2-\lambda|\leq L\},\quad \zeta\in\Z^2, L>0.$$

Using Chebyshev's inequality one can show that with probability $\gtrsim\frac{1}{N}$ we have
\begin{equation*}
\begin{split}
|\left\langle e_\zeta \psi_\lambda,\psi_\lambda\right\rangle|^2
\lesssim 
N\frac{\sum_{\xi\in A_0(\lambda,\lambda^\delta)}(|\xi-\zeta|^2-\lambda)^{-2}}
{\sum_{\xi\in A_0(\lambda,\lambda^\delta)}(|\xi|^2-\lambda)^{-2}}
\end{split}
\end{equation*}
where $\delta\in(\tfrac{\theta}{2},\tfrac{1}{2}-\theta)$ and $\theta=\frac{133}{416}$ is the best known exponent (due to Huxley \cite{Huxley}) in the circle law $$N(X)=\#\{|\xi|^2\leq X \mid \xi\in\Z^2\}=\pi X+O_\epsilon(X^{\theta+\epsilon}).$$

For a density one subsequence of new eigenvalues $\lambda$ we have that $\xi\in A_0(\lambda,\lambda^\delta)$ implies $$\xi\not\in A_\zeta(\lambda,\lambda^\delta) \Leftrightarrow ||\xi-\zeta|^2-\lambda|>\lambda^\delta$$ so that $$\frac{1}{||\xi-\zeta|^2-\lambda|}<\lambda^{-\delta}.$$
Also the error term in the circle law implies the following bound on the number of lattice points in a thin annulus $$\#A_0(\lambda,L)=O_\epsilon(\lambda^{\theta+\epsilon}).$$

Combining the bounds above we obtain the estimate (recall $\delta>\tfrac{1}{2}\theta$)
$$\sum_{||\xi|^2-\lambda|\leq \lambda^\delta} (|\xi-\zeta|^2-\lambda)^{-2}=O_\epsilon(\lambda^{-2\delta+\theta+\epsilon}).$$

Furthermore, we have for a density one subsequence of eigenvalues $\lambda$ the lower bound
$$\sum_{\xi\in A_0(\lambda,\lambda^\delta)}(|\xi|^2-\lambda)^{-2}\geq C(\epsilon)\lambda^{-\epsilon}.$$
So, for generic $\lambda$, the following bound holds with probability $\gtrsim \frac{1}{N}$ $$|\left\langle e_\zeta \psi_\lambda,\psi_\lambda\right\rangle|^2\lesssim_\epsilon N\lambda^{-1+3\theta+\epsilon}.$$

This argument can easily be extended for any trigonometric polynomial with nonzero frequencies. The result for a general torus $\T^2_L=\R^2/L\Z^2$ then follows by a simple scaling argument.

\section{Delocalization for random displacement models}

Let $B_L=[-L,L]^d$ and consider $$H_{\Omega,L}=-\Delta+\sum_{\xi\in\Z^d\cap B_L}\delta(x-\xi-\omega_\xi)$$ with Dirichlet boundary conditions, where the displacements $\omega_\xi$ are i.i.d. random variables with probability density $\PP\in C^0(\R^d)$ and $\supp\PP\subset B(0,\tfrac{1}{4})$.

Denote by $\psi_\lambda^L$ an $L^2$-normalized eigenfunction of $H_{\Omega,L}$. Fix $\chi\in C^\infty_c(\R^d)$, $\chi\geq0$ and $\|\chi\|_1=1$, and introduce the smoothed eigenfunction $$\Psi_\lambda^L(x)=\left(\int_{B_L}\chi(x'-x)|\psi_\lambda^L(x')|^2dx'\right)^{1/2}.$$

One of the problems with Definition \ref{loc} is that the localization centre depends on the sample $\omega\in\Omega$. To observe localization it is therefore advisable to consider the two-point correlation function of the smoothed eigenfunction $\Psi_\lambda^L$ which is defined for any two points $x,y\in B_L$ by
$$\Theta^L_\lambda(x,y)=\Psi_\lambda^L(x)\Psi^L_\lambda(y).$$ The decay is then observed with respect to the distance $|x-y|$, which is independent of the sample $\omega$.

We have the following alternative definition of localization.
\begin{defn}
Let $F\in C^0(\R_+)$ be strictly decreasing and $H_{\Omega,L}\psi_\lambda^L=\lambda\psi_\lambda^L$, $\|\psi_\lambda^L\|_2=1$.
We say that $H_{\Omega,L}$ is $F$-localized on an interval $I\subset\R_+$ if, for sufficiently large $L$, we have
$$\forall x,y\in B_L: \;\E\left(\sum_{\lambda\in I}\Theta_\lambda^L(x,y)\right)\leq F(|x-y|)$$
and the limit of the LHS as $L\to\infty$ exists.
\end{defn}

We can show that for large enough energy $H_{\Omega,L}$ fails to be localized. We denote the set of distinct eigenvalues of the Dirichlet Laplacian on $B_L$ by $\Lambda_L$.
\begin{thm}[H.U. 2015]
There exists $E_0\gg 1$ and a full density subsequence $\Lambda_L'\subset\Lambda_L$ such that for any $\lambda_k\in\Lambda_L'$, $\lambda_k>E_0$,
$$\E\left(\sum_{\lambda\in(\lambda_k,\lambda_{k+1})}\Theta^L_\lambda(x,y)\right)\gtrsim 1.$$
\end{thm}

In particular, this result implies that the localization length blows up at a certain critical energy threshold: $L_{loc}\rightarrow\infty$ as $E\to E_0$.

\subsection{The limit of large tori}

The key idea in proving the above theorem is to study the operator $H_{\Omega,L}$ in the limit $L\to\infty$. One of the key obstructions to doing this is the dependence on the number of potentials, $N\asymp L^d$, in the error term in equation \eqref{equidist}. 

However, we are able to improve the estimate of the error term significantly in the case of random displacement models: Fix $a_0\in C^\infty(B_1)$ and any $\epsilon>0$. Denote by $\Lambda=\{\lambda_j\}_{j=0}^\infty$ the set of distinct eigenvalues of the Dirichlet Laplacian on $B_1$. There exists a full density subsequence $\Lambda'\subset\Lambda$ such that we have, for sufficiently large $\lambda_k\in \Lambda'_L$, with probability $1-\epsilon$ for any $a\in C^\infty(B_1)$, $|\hat{a}|\leq|\hat{a_0}|$ and any $\lambda\in(\lambda_k,\lambda_{k+1})$
$$\int_{B_L}b(y)|\Psi_\lambda^L(y)|^2dy=\frac{1}{L^d}\left\{\int_{B_L}b(y)dy+O(\lambda^{-\delta_d}L^{-2\delta_d})\right\}, \quad b(\cdot)=L^{-d}a(\cdot/L).$$

Now let $a_0=\chi\in C^\infty(\R^d)$, $\|\chi\|_1=1$ and $\supp\chi\subset B(0,\epsilon_0)$ for some small $\epsilon_0$. We thus obtain the lower bound
$$\Psi_\lambda^L(x)\gtrsim L^{-d/2}$$ for any $L\gg 1$. In particular, we have the following lower bound for the two-point correlation function
$$\Theta_\lambda^L(x,y)\gtrsim |x-y|^{-d}, \quad|x-y|\asymp L.$$

In particular if we sum over all $\lambda\in\scrI=(\lambda_k,\lambda_{k+1})$, $\lambda_k\in\Lambda'_L$, and take the expectation, we may show
$$\E\left(\sum_{\lambda\in\scrI}\Theta^L_\lambda(x,y)\right)\gtrsim 1$$
where we have used that $$\#\{\lambda\in\scrI\}\asymp L^d.$$

\end{document}